\documentclass[nohyperref]{article}

\usepackage{microtype}
\usepackage{graphicx}
\usepackage{subfigure}
\usepackage{booktabs} 

\usepackage{hyperref}

\usepackage[accepted]{icml2022}

\usepackage{amsmath}
\usepackage{amssymb}
\usepackage{mathtools}
\usepackage{amsthm}
\usepackage{hyperref}

\usepackage[capitalize,noabbrev]{cleveref}

\theoremstyle{plain}

\theoremstyle{definition}

\theoremstyle{remark}

\usepackage[textsize=tiny]{todonotes}

\icmltitlerunning{Learnable wavelet neural networks for cosmological inference}

\begin{document}

\twocolumn[
\icmltitle{Learnable wavelet neural networks for cosmological inference}




\begin{icmlauthorlist}
\icmlauthor{Christian Pedersen}{a,b}
\icmlauthor{Michael Eickenberg}{c}
\icmlauthor{Shirley Ho}{a,b,d,e}
\end{icmlauthorlist}

\icmlaffiliation{a}{Department of Physics, New York University, New York City, USA}
\icmlaffiliation{b}{Centre for Computational Astrophysics, Flatiron Institute, New York City, USA}
\icmlaffiliation{c}{Centre for Computational Mathematics, Flatiron Institute, New York City, USA}
\icmlaffiliation{d}{Department of
Astrophysical Science, Princeton University, Peyton Hall, Princeton, USA}
\icmlaffiliation{e}{Department of Physics, Carnegie Mellon University, Pittsburgh, USA}

\icmlcorrespondingauthor{Christian Pedersen}{c.pedersen@nyu.edu}

\icmlkeywords{Machine Learning, ICML}

\vskip 0.2in
]

\printAffiliationsAndNotice{} 

\begin{abstract}
Convolutional neural networks (CNNs) have been shown to both extract more information than the traditional two-point statistics from cosmological fields, and marginalise over astrophysical effects extremely well. However, CNNs require large amounts of training data, which is potentially problematic in the domain of expensive cosmological simulations, and it is difficult to interpret the network. In this work we apply the learnable scattering transform, a kind of convolutional neural network that uses trainable wavelets as filters, to the problem of cosmological inference and marginalisation over astrophysical effects. We present two models based on the scattering transform, one constructed for performance, and one constructed for interpretability, and perform a comparison with a CNN. We find that scattering architectures are able to outperform a CNN, significantly in the case of small training data samples. Additionally we present a lightweight scattering network that is highly interpretable.
\end{abstract}

\section{Introduction}
\label{intro}
The process of extracting information from cosmological fields is a fundamental component of modern cosmology. The early Universe, observed as the Cosmic Microwave Background, is well described by a Gaussian random field, meaning the two-point function (or the power spectrum) contains all relevant information. However the growth of non-linear structure means this is not the case for probes of the late-time Universe, and it has been shown that significant information lies outside the two-point statistics in this regime. In the advent of several upcoming large surveys of late-time structure growth such as Euclid \cite{Euclid}, DESI \citep{DESI}, Roman observatory \citep{Roman}, Rubin observatory \citep{Rubin}, it is becoming increasingly important to consider methods of cosmological inference that go beyond the power spectrum, in order to maximise the scientific yield from these datasets.

The situation is further complicated by the fact that on small scales, astrophysical phenomena such as supernovae and AGN feedback affect the clustering of observational tracers. Recent work has demonstrated that neural networks are able to optimally marginalise over these effects \citep{Paco2020a}, and that convolutional neural networks (CNNs) are able to extract significantly more information from cosmological fields than the power spectrum \citep{Ravanbakhsh2016,Gupta2018,Paco2021a,Paco2021b,Lu2022} . However CNNs suffer from two potential pitfalls. First, the large numbers of parameters involved in deep convolutional networks require large amounts of training data to optimise. This is a significant consideration in the context of late-universe cosmology, where the hydrodynamical simulations required to model the non-linear baryon density field are extremely computationally expensive. Second, CNNs also suffer from a lack of interpretibility.

In this work we attempt to address these issues by presenting two models for cosmological inference based on the scattering transform \citep{Mallat2011}. Scattering transforms use a cascade of analytic wavelet transforms followed by complex modulus nonlinearities to construct descriptors of the input signal that behave smoothly to small deformations and that can be made to exhibit known symmetries of the data. Scattering transforms and the related wavelet phase harmonics have been successfully applied to cosmological parameter estimation from different types of input fields \cite{cheng2020new,allys2020new}. Recent work has demonstrated that learning the filter parameters can provide performance gains in the regime of small datasets \citep{Gauthier2021}. Using the suite of CAMELs simulations \citep{Paco2021b}, we investigate two kinds of scattering networks, one designed for performance and one designed for interpretability, and perform a comparison with a traditional CNN.

\begin{figure*}[ht]
\vskip 0.1in
\begin{center}
\centerline{\includegraphics[width=110mm]{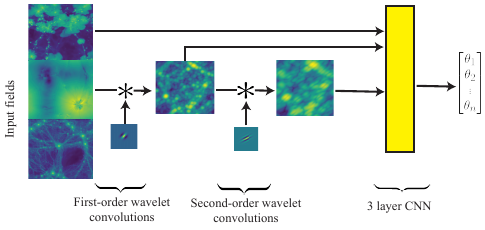}}
\caption{Architecture for the \textit{scattering network} (SN). The input fields are convolved with a set of 8 first order wavelet filters. These convolved fields are then convolved with a further set of 8 more second order filters. During each wavelet convolution the field is downsampled by a factor of 2. We include residual connections, so the zeroth and first order fields are smoothed and downsampled to match the resolution of the second order fields. These are then concatenated and passed to a 3 layer CNN, which provides the model output. The 4 parameters describing each wavelet filter are included in the backprop during training.}
\label{model1}
\end{center}
\vskip -0.3in
\end{figure*}

\section{Method}
\subsection{Simulations}
We use the publicly available CAMELs simulations \citep{Paco2021b}. Whilst we refer the reader to \cite{Paco2021b} for a full description, we briefly overview the most relevant aspects. We restrict our analysis to the subset of CAMELs simulations that use the IllustrisTNG simulation code \citep{Weinberger2017,Pillepich2018}. The dataset includes a suite of 1,000 simulations varying 6 parameters: $\Omega_m$ and $\sigma_8$ to describe cosmology, and $A_{\mathrm{SN}1}$, $A_{\mathrm{SN}2}$, $A_{\mathrm{AGN}1}$, $A_{\mathrm{AGN}2}$ that control the efficiency of the supernova and AGN feedback. Each simulation produces 13 different fields describing various properties such as gas temperature, total matter density, and magnetic fields. We use projected 2D slices of the 3D $25\times25\;h^{-3}\;\mathrm{Mpc}^3$ simulation boxes. Each slice is $5\;h^{-1}\;\mathrm{Mpc}$ thick, so with 5 slices along each axis, this produces a total set of 15,000 maps available for each field, each with a resolution of $256\times256$ pixels. These maps are then divided into training, test and validation sets at a fractional split of $0.9$, $0.05$ and $0.05$ respectively.

\subsection{Models}
We investigate three models \footnote{Code available on github at \href{https://github.com/Chris-Pedersen/LearnableWavelets}{github.com/Chris-Pedersen/LearnableWavelets}}. As our baseline CNN, we use a model with 12 convolutional layers, each with batch normalisation and Leaky ReLu activations.
Following \cite{Paco2022a} we train moment neural networks presented in \cite{Jeffrey2020} to infer the mean ($\mu_i$) and variance ($\sigma_i$) of the posterior for the 6 parameters ($\theta_i=\{\Omega_{m,i},\sigma_{8,i},A_{\mathrm{SN}1,i},A_{\mathrm{SN}2,i},A_{\mathrm{SN}1,i},A_{\mathrm{AGN}1,i},A_{\mathrm{AGN}2,i}\}$) describing each simulation. Therefore we define our loss function:
\begin{eqnarray}
\mathcal{L}&=&\sum_{i=1}^6\log\left(\sum_{j\in{\rm batch}}\left(\theta_{i,j} - \mu_{i,j}\right)^2\right)\nonumber \\
+&&\sum_{i=1}^6\log\left(\sum_{j\in{\rm batch}}\left(\left(\theta_{i,j} - \mu_{i,j}\right)^2 - \sigma_{i,j}^2 \right)^2\right)~
\label{eq:loss}
\end{eqnarray}

We also consider two models based on the scattering transform. First, we introduce the \textit{scattering network} (SN), shown in figure \ref{model1}. The first two layers are composed of convolutions with banks of 8 wavelet filters. We use Morlet wavelets, defined as
\begin{equation}\label{eq:morlet}
    \psi_{\sigma, \phi, \xi, \gamma}(u) = e^{-\|D_{\gamma}R_{\phi}(u)\|^2/(2\sigma^2)}(e^{i\xi u'}-\beta),
\end{equation}
where $\beta$ is a normalisation constant to ensure that the wavelet integrates to 0 over the spatial domain, $u' = u_1 \cos \phi + u_2 \sin \phi$, $R_{\phi}$ is the rotation matrix of angle $\phi$ and
$
D_{\gamma} = \begin{pmatrix}
1 & 0 \\
0 & \gamma
\end{pmatrix}
.
$
We therefore consider 4 parameters that modify the Morlet wavelet: the orientation is controlled by $\phi$, the spatial frequency by $\xi$, the Gaussian envelope is determined by $\sigma$, and the aspect ratio is set by $\gamma$. At initialisation, we randomly sample these parameters from $\phi \sim U[0,2\pi]$, $\xi \sim U[0.5,1]$, $\sigma \sim \log(U[\exp 1,\exp 5])$, and $\gamma \sim U[0.5,1.5]$. These filter parameters are
modified by gradient descent
as the network trains, and so are optimised to extract information from the fields they are operating on. Unlike in \citep{Gauthier2021}, we use different wavelet filters for first- and second-order scattering, allowing the network more flexibility to optimise these parameters.

\begin{figure*}[ht]
\vskip 0.2in
\begin{center}
\centerline{\includegraphics[width=120mm]{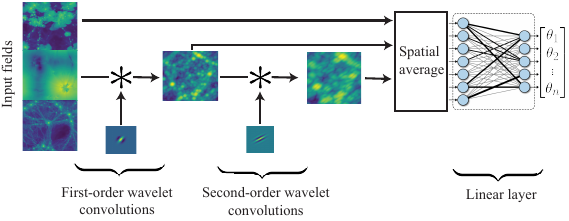}}
\caption{Model architecture for the \textit{interpretable network} (IN). The first two layers are identical to fig \ref{model1}, except now the output fields are spatially averaged, providing a single vector of 73 values. A single linear layer is used on this vector to predict the model output.}
\label{model2}
\end{center}
\vskip -0.3in
\end{figure*}

At each convolution, the input field is smoothed and downsampled by a factor of 2, so the second-order convolved fields have a resolution of $64\times64$. We include residual connections, so the zeroth and first-order fields are smoothed and downsampled to match this resolution and concatenated with the second order fields. This produces a total of $1+8+64=73$ fields output by the scattering layers. Finally these fields are passed to a shallow CNN with 3 convolutional layers, each with batch normalisation and ReLu activations. Finally we use one fully connected layer generating the output. We train this network on the same set of 11 fields as the ``MultiField" (MF) results from \citep{Paco2021b}, which includes all simulated fields except the dark matter mass and velocity fields.

Second, we also introduce an extremely lightweight model, the ``Interpretable network" (IN) shown in figure \ref{model2}. The first two layers of this network are identical to the SN, except now the 73 output fields of the scattering layers are spatially averaged to form a single vector of 73 numbers. We then use a single linear layer on this vector to obtain parameter values. For the sake of simplicity in interpreting this network, we train this model on only the cold dark matter mass field ($M_\mathrm{cdm}$) to produce the results in section \ref{ss:IN}.

For all models, we use batch gradient descent with Adam optimiser and a batch size of $32$. Hyperparameters are optimised using the \texttt{optuna} package using $30$ trials. We vary independently the learning rate for the scattering and neural layers, and in the case of the SN and CNN, we vary the number of convolutional filters.

\section{Results}

\begin{table}[t]
\vskip 0.15in
\begin{center}
\begin{small}
\begin{sc}
\begin{tabular}{lcccr}
\toprule
Model & $\Delta \Omega_m (\%)$ & $\Delta \sigma_8$ (\%) & Valid. Loss\\ 
\midrule

CNN (10k)             & 6.31 & 4.02 & -11.90    \\
SN (10k)              & 4.24 & 3.42 & -12.18    \\
\midrule    
CNN (5k)              & 6.61 & 5.15 & -11.18    \\
SN (5k)               & 5.29 & 4.15 & -11.58    \\
\midrule
CNN (1k)              & 10.07 & 8.99  & -9.82     \\
SN (1k)               & 6.32  & 4.33  & -10.62    \\
\midrule
SN (1k $M_{\mathrm{cdm}}$)     & 2.48 &  1.84  & -3.83   \\
IN (1k $M_{\mathrm{cdm}}$)     & 11.50 &  2.79 & -3.33   \\
\bottomrule
\end{tabular}
\end{sc}
\end{small}
\end{center}
\caption{Comparison of model performance for datasets of sizes 10,000, 5,000 and 1000 simulated maps. We show results for the empirical accuracy on $\Omega_m$ and $\sigma_8$ averaging over the test set, and the validation loss. For the final two rows, we use only the $M_{\mathrm{cdm}}$ fields. For simplicity, for these last examples we only model the mean of the posterior, i.e. the loss function is only the first term in equation \ref{eq:loss}, which is the reason for the large difference in stated values.}\label{tab:performance}
\vskip -0.4in
\end{table}

\subsection{Scattering network}
We show a comparison of model performance between the baseline CNN and SN in table \ref{tab:performance}. We focus on the accuracy of the predicted values for $\Omega_m$ and $\sigma_8$, when averaging over the test set. We also show the validation loss as evaluated by equation \ref{eq:loss}. The two models are compared for 3 different dataset sizes: $10,000$, $5,000$ and $1,000$, in order to evaluate how performance scales with the size of the training set. These numbers include test and validation sets. We find significantly better performance from the SN when using smaller datasets. This is consistent with a similar comparison performed in \citep{Gauthier2021} on the CIFAR-10 dataset. The SN model has approximately an order of magnitude less parameters than the CNN, with $\sim 2\times10^6$ versus $2\times 10^7$, and is therefore able to optimise effectively on smaller datasets. In addition, our results indicate that even as we scale to larger datasets, the SN performance mildly outperforms the CNN.

\begin{figure*}[ht]
\vskip 0.2in
\begin{center}
\centerline{\includegraphics[width=150mm]{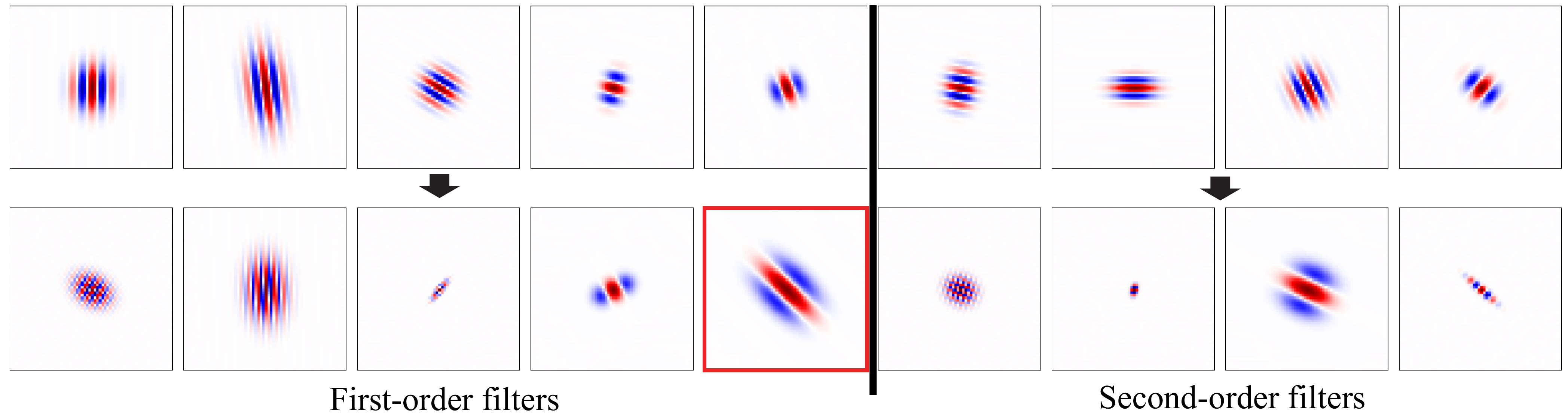}}
\caption{Wavelet filters before (top) and after (bottom) training. We show only the 9 most relevant fields after applying $L_1$-penalised linear regression to the IN described in figure \ref{model2}. We find that these 9 filters plus the zeroth order filters are able to retain $98\%$ of the model performance. All second order fields selected by the network were the product of a single first order wavelet, highlighted in red. For visualisation purposes we have magnified the spatial size of the wavelets by a factor of 4 in this figure.}
\label{wavelets}
\end{center}
\vskip -0.2in
\end{figure*}

\subsection{Interpretable network}
\label{ss:IN}

\begin{figure}[ht]
\vskip 0.2in
\begin{center}
\centerline{\includegraphics[width=90mm]{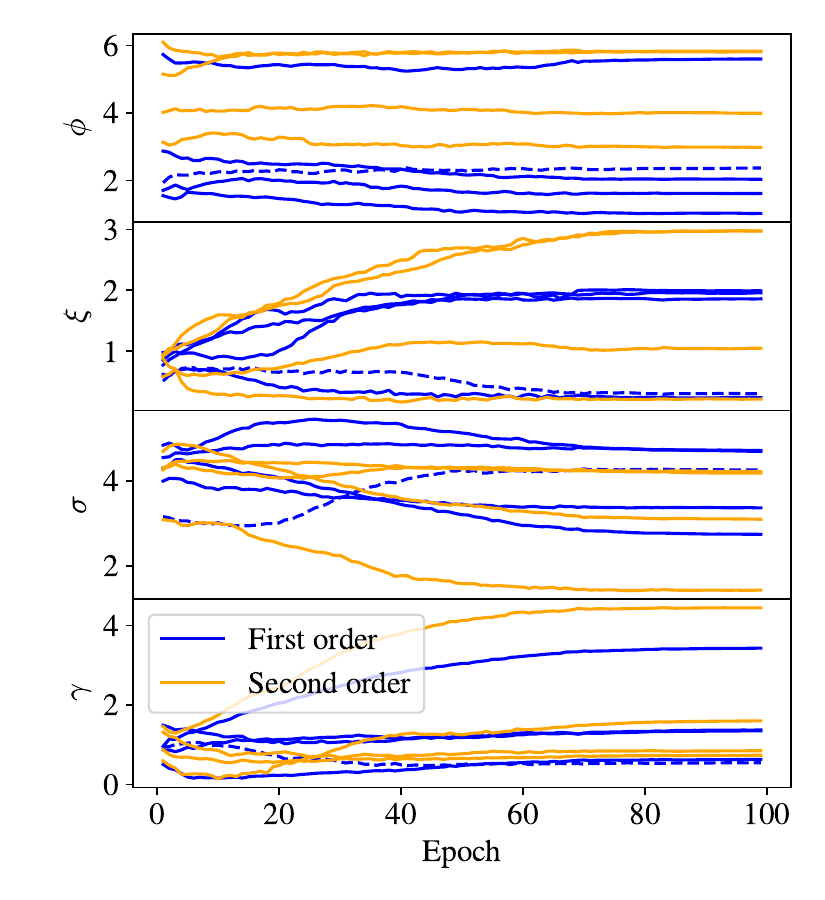}}
\caption{Evolution of the wavelet parameters describing each filter shown in figure \ref{wavelets} during training. First order wavelets are shown in blue and second order wavelets are shown in orange. The first order filter used to propagate fields to the second order filters is shown in blue dashed lines.}
\label{wavelet_params}
\end{center}
\vskip -0.2in
\end{figure}

At the bottom of table \ref{tab:performance} we compare the performance of the SN and IN when using a set of $1,000$ maps of the $M_\mathrm{cdm}$ field. First, we find that the SN is able to predict $\Omega_m$ and $\sigma_8$ to a remarkable degree of accuracy when using only $1,000$ maps. This is potentially due to the fact that when operating on only one type of input field, the wavelet parameters are able to better 
extract information from 
this field. Whilst the difference in performance on $\Omega_m$ is very large, the IN is able to predict $\sigma_8$ to $2.79\%$ accuracy despite having only $654$ parameters.

Finally, 
after training the filters for the IN, 
we use $L_1$-penalised linear regression to
isolate the most significant filters in the model. We find that using only 10 of the 73 elements of the final linear layer, we are able to retain $98\%$ of the model performance of the IN. These 10 fields correspond to the zeroth order field, 5 first-order wavelets, and 4 second-order wavelets. We show these 9 wavelets before and after training in figure \ref{wavelets}. Interestingly, all 4 of the second-order fields are downstream from the same first-order wavelet, which is highlighted in red. It is possible that the network has chosen to optimise this filter 
to operate in conjunction with the second order fields.

In figure \ref{wavelet_params}, we show the evolution of the parameters describing these 9 wavelet filters during training of the IN. Whilst the orientations of the wavelets do not change significantly, the frequency and aspect ratios are clearly strongly driven by the input fields to different values than they were initialised with.

\section{Conclusion}
We performed a comparison between a learnable scattering transform and a CNN for the purposes of inferring cosmological parameters and marginalising over astrophysical effects. We find a scattering network is able to outperform a CNN, with the performance difference significantly increasing for smaller training set sizes. We also present a lightweight interpretable scattering network that is able to find a sparse wavelet compression of the input fields.

\clearpage
\bibliography{example_paper}
\bibliographystyle{icml2022}

\end{document}